\documentclass{IEEEoj}
\usepackage{cite}
\usepackage{amsmath,amssymb,amsfonts}
\usepackage[noend]{algorithmic}
\usepackage{graphicx,color}
\usepackage{textcomp}
\usepackage[acronym, nolist]{glossaries}
\usepackage{underscore}
\usepackage{algorithm}
\usepackage{orcidlink}
\usepackage[normalem]{ulem}
\hypersetup{
	colorlinks=true, 
	linkcolor=blue!70!black, 
	citecolor=red!70!black, 
	filecolor=blue!70!black, 
	urlcolor=blue!70!black, 
}

\def\BibTeX{{\rm B\kern-.05em{\sc i\kern-.025em b}\kern-.08em
    T\kern-.1667em\lower.7ex\hbox{E}\kern-.125emX}}
\begin{document}
\receiveddate{XX Month, XXXX}
\reviseddate{XX Month, XXXX}
\accepteddate{XX Month, XXXX}
\publisheddate{XX Month, XXXX}
\currentdate{XX Month, XXXX}

\newacronym{MoM}{MoM}{method-of-moments}
\newacronym{EM}{EM}{electromagnetic}
\newacronym{NFL}{NFL}{no free lunch}
\newacronym{ML}{ML}{machine learning}
\newacronym{AI}{AI}{artificial intelligence}
\newacronym{NGnet}{NGnet}{normalized Gaussian network}
\newacronym{TO}{TO}{topology optimization}
\newacronym{LLM}{LLM}{large language model}
\newacronym{MMA}{MMA}{method of moving asymptotes}
\newacronym{SGD}{SGD}{stochastic gradient descent}
\newacronym{GD}{GD}{gradient descent}
\newacronym{EFIE}{EFIE}{electrical field integral equation}
\newacronym{DOF}{DOF}{degrees of freedom}
\newacronym{RWG}{RWG}{Rao–Wilton–Glisson}
\newacronym{PEC}{PEC}{perfect electric conductor}
\newacronym{AD}{AD}{automatic differentiation}
\newacronym{PINNs}{PINNs}{physics-informed neural networks}
\newacronym{BO}{BO}{Bayesian optimization}
\newacronym{RL}{RL}{return loss}
\newacronym{FBW}{FBW}{fractional bandwidth}
\newacronym{TM}{TM}{transverse magnetic}
\newacronym{TE}{TE}{transverse electric}

\newcommand{\norm}[1]{\left\lVert#1\right\rVert_{2}}
\newcommand{\ovarb}{\V{\T{g}}}
\newcommand{\ovar}{\V{\T{g}}_\T{c}}
\newcommand{\ovart}{\V{\widetilde{\T{g}}}_\T{c}}
\newcommand{\ovarn}{\T{g}_{\T{c},{n}}}
\newcommand{\resistivity}{r_n}
\newcommand{\resistivitym}{\V{r}}
\newcommand{\Loss}{\mathcal{L}}
\newcommand{\projMatrix}{\V{\Theta}}
\newcommand{\projMatrixn}{\Theta_{nn}}
\newcommand{\denstityKernel}{\V{H}}
\newcommand{\denstityKerneln}{H_{nm}}
\newcommand{\Lbo}{\mathcal{L}_{\T{BO}}}
\newcommand{\Qu}{Q_\T{U}}
\newcommand{\Qe}{Q_\T{E}}
\newcommand{\Qz}{Q_\T{z}}
\newcommand{\Qb}{Q_\T{B}}
\newcommand{\Qlb}{Q_\T{lb}^\T{TM}}
\newcommand{\w}{w(\ovarn)}
\newcommand{\bigO}{\mathcal{O}}
\newcommand{\pn}{p_\T{N}}
\renewcommand{\algorithmiccomment}[1]{\bgroup\hfill $\triangleright$ ~#1\egroup}
\providecommand{\V}[1]{\boldsymbol{#1}}
\providecommand{\M}[1]{\mathbf{#1}} 
\providecommand{\herm}{\mathrm{H}}
\providecommand{\Ivec}{\ensuremath{\M{I}}}
\providecommand{\Vvec}{\ensuremath{\M{V}}}
\providecommand{\Zmat}{\ensuremath{\M{Z}}}
\providecommand{\Ymat}{\ensuremath{\M{Y}}}
\providecommand{\Rmat}{\ensuremath{\M{R}}}
\providecommand{\Xmat}{\ensuremath{\M{X}}}
\providecommand{\Wmat}{\ensuremath{\M{W}}}
\providecommand{\srcRegion}{\ensuremath{\varOmega}} 
\providecommand{\Quot}[1]{``{#1}''} 
\providecommand{\T}[1]{\mathrm{#1}}
\providecommand{\Prad}{P_\T{rad}}
\providecommand{\We}{\ensuremath{W_\mathrm{e}}}
\providecommand{\Wm}{\ensuremath{W_\mathrm{m}}}
\providecommand{\XEmat}{\ensuremath{\M{X}_\T{e}}}
\providecommand{\XMmat}{\ensuremath{\M{X}_\T{m}}}
\providecommand{\Rin}{\ensuremath{R_\T{in}}}
\providecommand{\Xin}{\ensuremath{X_\T{in}}}
\providecommand{\Zin}{\ensuremath{Z_\T{in}}}
\providecommand{\J}{\ensuremath{\mathrm{j}}}    
\providecommand{\RE}{\ensuremath{\mathrm{Re}}}  
\providecommand{\IM}{\ensuremath{\mathrm{Im}}}  

\title{Antenna Q-Factor Topology Optimization with Auxiliary Edge Resistivities}

\author{\authorrefmark{1}Stepan Bosak\,\orcidlink{0009-0007-3557-2155}, \authorrefmark{1}Miloslav Capek\,\orcidlink{0000-0002-7442-889X} (Senior Member, IEEE), \authorrefmark{2}Jiri Matas\,\orcidlink{0000-0003-0863-4844}}
\affil{Department of Electromagnetic field}
\affil{Department of Cybernetics\\
       Faculty of Electrical Engineering, Czech Technical University in Prague, Czech Republic}
\corresp{CORRESPONDING AUTHOR: Stepan Bosak (\texttt{stepan.bosak@fel.cvut.cz}).}
\authornote{This work was supported by The Czech Science Foundation under project~\mbox{No.~21-19025M} and by the Czech Technical University in Prague under project Advanced methods for modelling electromagnetic structures:~\mbox{SGS25/143/OHK3/3T/13.} }
 
\begin{abstract}
This paper presents a novel bi-level topology optimization strategy within the method-of-moments paradigm. The proposed approach utilizes an auxiliary variables called edge resistivities related to the Rao–Wilton–Glisson method-of-moments basis functions, for a definition of a fast local optimization algorithm. The local algorithm combines automatic differentiation with adaptive gradient descent. A Bayesian optimization scheme is applied on top of the local algorithm to search for an optimum position of the delta-gap feeding and optimizer hyperparameters. The strength of the algorithm is demonstrated on Q-factor minimization for electrically small antennas. Auxiliary edge resistivity topology optimization outperforms current state-of-the-art topology optimization methods, including material density-based approaches and memetic schemes, in terms of convergence. However, due to the nature of gradient descent, careful tuning of the optimizer hyperparameters is required. Furthermore, the proposed method solves the known binarization issue. Two designs that achieved self-resonance and approached the Q-factor lower bound were further assessed in CST Microwave Studio. 
\end{abstract}


\maketitle

\section{Introduction}
\IEEEPARstart{T}{he} common approach to antenna design starts by picking a suitable textbook solution~\cite{BalanisModernAntennaHandbook}, followed by a refinement guided by the engineer's intuition, experience, and skills towards a desired objective. However, if none of the obtained antennas meet the required performance, the procedure needs a restart, increasing the duration of the design and its cost. Automated design changes this paradigm.

Automated design has recently emerged as a viable alternative to hand-crafted approaches~\cite{2012PixelingRahmat}\cite{MLAO-survey}\cite{topoOpt}. 
In this context, automated design typically refers to \gls{TO}, a computational technique that systematically determines the best material layout for a given performance goal. \gls{TO} may produce highly novel and non-intuitive material distributions and can perform incremental improvements of a known design~\cite{2021capeketalTSGAmemeticsPart2}. Throughout this work, we use \Quot{automated design} and \Quot{topology optimization} interchangeably.

A related term, inverse design, is often used in the fields of photonics and electromagnetics to describe the same concept: specifying the desired performance and allowing an algorithm to find a structure that achieves it. For a detailed overview of inverse design methods in photonics, refer to~\cite{Molesky2018}.

\gls{TO} methods fall into two categories: global~\cite{heuristics} and local~\cite{2016LiuAMS}. In recent publications, these methods are combined to improve robustness or convergence speed~\cite{AI-DE}, which are contradictory attributes according to the \gls{NFL} optimization theorem~\cite{NFL}. 

Global \gls{TO}~\cite{2012PixelingRahmat}, most commonly represented by evolutionary algorithms~\cite{heuristics} or particle swarm optimization~\cite{2012PixelingRahmat}, omits the need to compute gradients and approaches the optimized device as a black box. These methods apply heuristic algorithms and are robust. However, they do not provide guarantees with respect to global optimality. Moreover, the number of evaluations necessary to achieve the desired performance increases dramatically with the number of optimized variables~\cite{1999JohnsonGAMoM}.

Local \gls{TO} typically exploits either gradient analysis~\cite{topoOpt} or discrete differences~\cite{2021capeketalTSGAmemeticsPart1}. Local \gls{TO} converges fast~\cite{2011SigmundGA}, and based on the optimization problem and formulation, enables optimization of billions of design variables~\cite{2020BaandruptopoBridge}. It has been shown that when gradient information is available, convergence is faster than with global \gls{TO}~\cite{2011SigmundGA}. However, these methods are often sensitive to initial conditions; therefore, they require additional tuning and knowledge of the underlying problem. 

Recently, increasing attention has been given to \gls{TO} approaches incorporating \gls{ML}, see extensive reviews~\cite{AI-survey, 2022WoldsethToAI} and references therein. \gls{ML}-assisted optimization has been applied in both global and local \gls{TO} frameworks. In global \gls{TO}, \gls{ML} is commonly employed as a surrogate model~\cite{Koziel2014AntennaDB}, predicting the performance of candidate designs and thereby accelerating the optimization process~\cite{MLAO}. System-by-Design~\cite{2022MassaSbD} is yet another emerging paradigm in global optimization for designing \gls{EM} devices based on surrogate models and efficient sampling strategies. It is closely related to \gls{BO}~\cite{1998JonesBO}.

\gls{BO} is a black-box heuristic optimization technique commonly used to optimize hyperparameters in neural networks~\cite{2018FrazierBO}. It relies on a surrogate model, typically a Gaussian process. Although \gls{BO} does not guarantee convergence to the global minimum, there is empirical evidence for its ability to find high-quality solutions~\cite{1998JonesBO}.

An alternative global method (not a surrogate model), known as \gls{NGnet}~\cite{NGNet-s}, represents the distribution of the material as a weighted sum of normalized Gaussians~\cite{1989MoodyFLN}.

Another line of research explores \gls{PINNs}, which embed the governing physical equations in the architecture of the neural network~\cite{2019RaissiPINNs}. \gls{PINNs} have been applied to both forward modeling and inverse design tasks, with applications in global~\cite{2019RaissiPINNs} and local \gls{TO} methods~\cite{Jeong2023PINNTO}.
 
\Gls{MoM}~\cite{MOM} as a computation tool allows a fast algebraic formulation of \gls{TO}.
\gls{MoM}-based \gls{TO} has received a great deal of attention~\cite{1999JohnsonGAMoM}\cite{2009ToivanenADMoM}\cite{2016LiuAMS}\cite{2021capeketalTSGAmemeticsPart1}. However, even novel gradient-based \gls{TO} within the \gls{MoM} paradigm~\cite{Q-TOPO} struggles to surpass the handcrafted designs~\cite{2019CapekOA}, and a large number of \gls{DOF} (tens of thousands) needs to be utilized for these methods to be competitive. 
Unlike gradient-based methods, state-of-the-art \gls{MoM}-based global methods usually surpass handcrafted designs~\cite{2021capeketalTSGAmemeticsPart2}, but the convergence time rapidly increases with the number of \gls{DOF}, and often the resulting shapes are irregular and difficult to manufacture~\cite{2022PGSCapek}. 

This paper presents a \gls{MoM} \gls{RWG} method that connects a gradient-based and a global algorithm. The local method utilizes auxiliary edge resistivities, variables that allow gradient propagation with respect to the addition or removal of \gls{RWG} basis functions~\cite{CapeketalShapeSynthesisBasedOnTopologySensitivity}. This novel formulation outperforms state-of-the-art local algorithms in terms of the self-resonant Q-factor~\cite{YaghjianBestImpedanceBandwidthAndQOfAntennas}, \cite{2025YaghjianQFBW} for electrically small antennas in rectangular design region.

We employ \gls{AD}~\cite{2018BaydinAD} to compute the gradients and optimize using a variant of \gls{GD}~\cite{2017LoshchilovSGDR}. Lastly, \gls{BO}~\cite{2018FrazierBO} searches for optimizer hyperparameters and the position of delta-gap feeding~\cite{balanis2012advanced}.

This paper is structured as follows. Section~\ref{sec:MoMTopo} explains the definition of auxiliary edge resistivities within the \gls{EFIE} \gls{MoM} \gls{RWG} paradigm. Section~\ref{sec:Qfactor} briefly introduces the antenna Q-factor and the definition of the minimization objective. Section~\ref{sec:routine} gives details on the optimization algorithm. Section~\ref{sec:results} digests the numerical investigations. Lastly, Section~\ref{sec:conclusions} summarizes the article. The code for the auxiliary edge resistivity \gls{TO} and the bi-level optimization routine for minimization of the Q-factor is available in~\cite{2025BosakAERGitHub}.  

\section{Auxiliary Edge Resistivity MoM Topology Optimization}
\label{sec:MoMTopo}
This section outlines the numerical modeling based on \gls{MoM}~\cite{MOM} and commonly applied \gls{RWG} basis functions~\cite{1982RWGRao}. Special attention is paid to the inclusion of auxiliary edge resistivities, which play a key role in the optimization framework.

\subsection{EFIE Matrix Form}
For an arbitrary-shaped radiator~$\srcRegion$ made of a good conductor, the \gls{EFIE}~\cite{HarringtonTimeHarmonicElmagField} relates the current density flowing on the radiator surface and the tangential component of the total electric field. To find an unknown current density,~$\srcRegion$ is discretized into surface elements over which suitable basis functions are defined. In this work, the \gls{RWG} basis functions~\cite{1982RWGRao}, defined on pairs of adjacent triangular elements, are used, see Fig.~\ref{fig:rwgrelaxation}a). The resulting system of equations is then solved using the \gls{MoM}~\cite{MOM}.

The Galerkin procedure~\cite{ChewTongHuIntegralEquationMethodsForElectromagneticAndElasticWaves} recasts the \gls{EFIE} into a matrix form
\begin{equation}
\label{eq:mom}
\Zmat\Ivec = \Vvec\rightarrow\Ivec = \Zmat^{-1}\Vvec = \Ymat\Vvec,
\end{equation}
where~$N$ is the number of basis functions~$\psi_n$, $\Zmat\in \mathbb{C}^{N\times N}$ is the vacuum impedance matrix, $\Ymat\in\mathbb{C}^{N\times N}$ is the admittance matrix, $\Ivec\in \mathbb{C}^{N\times 1}$ is a vector of expansion coefficients for the currents, and ${\Vvec\in \mathbb{C}^{N\times 1}}$ is a vector of excitation coefficients~\cite{MOM}. The position and magnitude of excitation are known prior to solving~\eqref{eq:mom}. In this work, we assume a single-port antenna, for which Q-factor is well defined~\cite{YaghjianBestImpedanceBandwidthAndQOfAntennas}. Specifically, and without any loss of generality, the antenna is fed by a discrete delta-gap feeder~\cite{balanis2012advanced}. Equation~\eqref{eq:mom} implicitly defines a relation between the shape of the structure through the admittance matrix, feeding, and induced currents. All subsequent operations, for example, calculation of the Q-factor, are performed over this discretized system.

\subsection{Definition of Auxiliary Edge Resistivities}
\label{subsubsec:aer}
Auxiliary edge resistivities are variables associated with the \gls{RWG} basis functions. A previous study~\cite{CapeketalShapeSynthesisBasedOnTopologySensitivity} reaffirmed that the removal of a basis function~$\boldsymbol{\psi}_n$, corresponds to introducing a lumped resistor with infinite resistance. The authors demonstrated that such removal is equivalent to carving a slot into the radiating surface. This finding is extended here by introducing the intermediate resistivities of these lumped resistors, auxiliary edge resistivities. 

\begin{figure}
\centerline{\includegraphics[width=\linewidth]{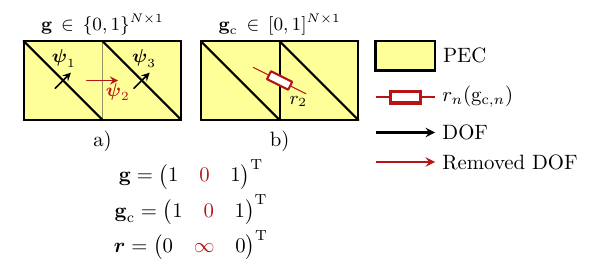}}
\caption{A diagram of basis function removal. a) Basis function~$\boldsymbol{\psi}_2$ is removed. b) The removal of~$\boldsymbol{\psi}_2$ is equivalent to set~$r_2 = \infty$. Vector~$\ovarb$ is a binary indicator denoting the presence~($1$) or absence~($0$) of a DOF. Vector~$\ovar$ is a relaxed version of~$\ovarb$, allowing~$\boldsymbol{r}$ to take intermediate resistivity values instead of being restricted to~$0$ or~$\infty$.}
\label{fig:rwgrelaxation}
\end{figure}

Let us define a removal/addition of a basis function~$\boldsymbol{\psi}_n$ using a binary vector~$\ovarb \in \{0, 1\}^{N \times 1}$, where~$0$ indicates absence and~$1$ indicates presence. Fig.~\ref{fig:rwgrelaxation} shows an example of the removal of a basis function~$\boldsymbol{\psi}_2$, which can be interpreted as~$r_2 = \infty$. Subsequently, we relax the binary vector~$\ovarb$ into its continuous form~$\ovar \in [0, 1]^{N\times 1}$, which means that resistivities~$\resistivity$ no longer need to be of value~$0$ or~$\infty$. The lumped elements procedure allows for a modification of the \gls{MoM} state equation~\eqref{eq:mom} using a diagonal lumped element matrix first introduced in~\cite{Harrington1972}. The new state equation reads
\begin{equation}
    \label{eq:momrelaxed}
    [\Zmat + \projMatrix(\ovar)]\Ivec = \Vvec,
\end{equation}
where~$\projMatrix(\ovar) \in \mathbb{R}^{N\times N}$ defines the auxiliary resistivity assigned to basis functions~$\V{\psi}$, and controls the current density between two adjacent triangles.

The advantage of auxiliary edge resistivities, compared to the conventional material approach defined in~\cite{Q-TOPO}, is that one \gls{DOF} represents only one row and one column in the system of linear equations~\eqref{eq:momrelaxed} instead of multiple rows and columns~\cite{Q-TOPO}. For the same number of \gls{DOF} of both methods, this property significantly reduces the computational burden, considering that the complexity of solving~\eqref{eq:momrelaxed} is $\bigO(N^3)$. However, the disadvantage of the approach is that the method can generate patches of material isolated by infinitely thin slots~\cite{2025FriendlyShapedNeuman}, especially at the boundaries between the void and the material. 

\subsection{Resistivity Interpolation}
\label{subsec:resproj}
The objective of optimization is to remove or add pieces of material in the optimization region~$\srcRegion$. Removal or addition is achieved by electrically shortening or isolating pieces of material using the lumped element resistivities~$\resistivity$. The interpolation between the design variable~$\ovarn$ and the physical resistivity~$\resistivity$ is defined according to~\cite{Erentok2011SubWavelengthAntennas} as
\begin{equation}
    \label{eq:resproj}
    \resistivity(\ovarn) = R_\T{o}\left(\frac{R_\T{s}}{R_\T{o}}\right)^{f(\ovarn)},
\end{equation}
where function~$f(\ovarn)$ aggregates penalization function~\cite{topoOpt}, density convolution filter~\cite{2001BrunselasticStructures}, and projection function replacing hard thresholding~\cite{2010projectionWang}.

Finally the diagonal matrix~$\projMatrix(\ovar)$ is defined element-wise as~\cite{CapeketalShapeSynthesisBasedOnTopologySensitivity}
\begin{equation}
    \projMatrixn = \resistivity l_n^{2},
\end{equation}
where $\resistivity$ is the previously evaluated resistivity based on~\eqref{eq:resproj}, and $l_n$ is the respective edge length.

\subsection{Projection Methods}
\label{subsec:projection}
The projection functions~$f$ used in interpolation function~\eqref{eq:resproj} are a standard composition of functions utilized in \gls{TO}~\cite{topoOpt}, including the penalization function~\cite{topoOpt}, the density convolution filter~\cite{2001BrunselasticStructures}, and the thresholding function~\cite{2010projectionWang}. Details about these methods are included in the Appendix~\ref{Appendix:proj}.

\begin{figure}
\centerline{\includegraphics[width=\linewidth]{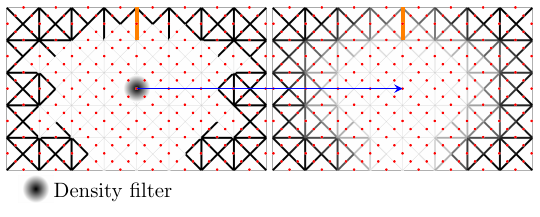}}
\caption{Left -- thresholded shape with edge centers in red. The density filter is applied on values~$\ovar$ based on the second norm between the center point~$c_m$ and surrounding points~$c_n$. Right -- the result of applying density filtering on $\ovar$. }
\label{fig:filter}
\end{figure}

The density filter~\cite{2001BrunselasticStructures} is modified for auxiliary edge resistivities (see Appendix~\ref{Appendix:proj}), and its effect is shown in Fig.~\ref{fig:filter}. The left panel shows a black and white structure. The density filter is applied to each point~$c_m$ representing the center of the filter and the center of the corresponding edge and neighboring points~$c_n$ based on the second norm. The right panel represents the result of the filtering operation. Furthermore, the delta-gap feeder~\cite{balanis2012advanced} indicated in orange remains unchanged after filtering. It should be noted that the filtering approach works for any mesh, regardless of the electromagnetic excitation.

\section{Q-factor Minimization}
\label{sec:Qfactor}
For self-resonant~\cite{2017CapekQmin} electrically small antennas ($ka < 1$, where~$k$ is the wave number and~$a$ the radius of the smallest circumscribed sphere), the Q-factor provides a measure of the antenna \gls{FBW} as $\T{FBW}\propto Q^{-1}$~\cite{2025YaghjianQFBW}. The standard definition of Q-factor is~\cite{IEEEStdantennas}
\begin{equation}
    \label{eq:q}
    Q(\Ivec) = \frac{2\omega\max\{\We,\Wm\}}{\Prad},
\end{equation}
where~$\We$ and~$\Wm$ represent stored electric and magnetic energies~\cite{Vandenbosch2010} and $\Prad$ stands for the radiated power. For further details on the Q-factor, see the Appendix~\ref{Appendix:q}.

A self-resonant Q-factor is achieved when $\Wm - \We = 0$~\cite{2016capekqModesOptim}. A general Q-minimization problem can be defined as
\begin{equation}
\begin{aligned}
    \label{eq:generalq}
    \min \quad & Q \\
    \textrm{s.t.} \quad & \Wm - \We = 0 .
\end{aligned}
\end{equation}
Minimizing the Q-factor~\eqref{eq:q} without additional constraints should result in a self-resonant antenna~\cite{2016capekqModesOptim}, but previous research has observed that this wasn't always necessarily the case~\cite{Q-TOPO} because of imperfections of the optimizer and limited resolution of the underlying disretization\footnote{Having a sufficient discretization, matching can always be done by a 3D structure being a part of the antenna layout, e.g., an inter-digital capacitor which would replace the external lumped matching element.}.

\subsection{Q-factor Minimization via Auxiliary Edge Resistivities}
In this work, we utilize a scalarized form of the optimization problem~\eqref{eq:generalq} defined as~\cite{2021capeketalTSGAmemeticsPart1}
\begin{equation}
    \label{eq:opimization}
    \begin{aligned}
        \min_{\ovar} \quad & \Loss(\ovar) = \frac{Q(\ovar) + \delta \Qe(\ovar)}{\Qlb} + \gamma w(\ovar)\\
        \textrm{s.t.} \quad & \Ivec = [\Zmat + \projMatrix(\ovar)]^{-1}\Vvec \\
        & 0 \leq \ovarn \leq 1, \quad n=1\dots N,   \\
    \end{aligned}
\end{equation}
where $Q$ is the standard Q-factor defined in~\eqref{eq:q} and rewritten (see Appendix~\ref{Appendix:q})
\begin{equation}
    \label{eq:qu}
    Q(\ovar) = \frac{\max\{\Ivec^{\herm}\XEmat\Ivec, \Ivec^{\herm}\XMmat\Ivec\}}{\Ivec^{\herm}\Rmat\Ivec},
\end{equation}
and~$\Qe$ is a self-resonance tuning term~\cite{2016capekqModesOptim}
\begin{equation}
    \label{eq:qe}
    \Qe(\ovar) = \frac{|\Ivec^{\herm}\Xmat\Ivec|}{\Ivec^{\herm}\Rmat\Ivec} = \frac{|\Ivec^{\herm}(\XMmat-\XEmat)\Ivec|}{\Ivec^{\herm}\Rmat\Ivec},
\end{equation}
which measures the amount of stored energy needed to achieve a self-resonant Q-factor. The auxiliary edge resistivities are represented via the optimization variable~$\ovar$. The current coefficients~$\Ivec$ are evaluated using $\ovar$ and the Q-factors are calculated using~$\Ivec$. The matrices~$\Rmat$ and~$\Xmat$ represent the real and imaginary parts of the \gls{EFIE} vacuum impedance matrix respectively. Matrices~$\XEmat$~and~$\XMmat$ represent the electric and magnetic reactances.
The function~$\w$ is the binary regularization function defined as~\cite[(41)]{2007SigmundMOTopo}
\begin{equation}
    \label{eq:regularization}
    \w = \sum\limits_{n} \frac{4\ovarn(1 - \ovarn)}{N},
\end{equation}
where~$N$ is the number of \gls{RWG} basis functions. This function eliminates the effect of hard thresholding on performance~\cite{Q-TOPO}.
Coefficients~$\delta$ and~$\gamma$ are penalization coefficients used throughout the optimization to push the optimizer toward the self-resonant Q-factor and binary material distribution, respectively, and represent two optimization meta-parameters that the designer should choose.  

This paper compares all the results with the lower bound of the electric dipole type Q-factor (\gls{TM} mode Q-factor)~$\Qlb$, which is obtained from the expansion of the radiated field into spherical modes and the division of the radiated power (hence the matrix~$\Rmat$) into \gls{TM} and \gls{TE} parts~\cite{2015GustafssonPBA}. Q-factor \gls{TM} lower-bound is a tight bound for an electrically small single-port antenna~modes~\cite{2019CapekOA}, and its value~$\Qlb$ is obtained by optimizing the current expansion coefficients~$\Ivec$ in the \gls{MoM} system~\cite{2017CapekQmin}.

\section{Optimization Routine}
\label{sec:routine}
The proposed optimization routine solves~\eqref{eq:opimization} by applying \gls{AD}~\cite{2018BaydinAD} and using a \gls{GD}-based optimization algorithm called AdamW~\cite{2017LoshchilovAdamW}.  

\gls{AD} (in the backward mode) computes the forward process and records the intermediate dependencies between the input and output in a computational graph. In the second step, derivatives are calculated by propagating adjoints in reverse from output to input~\cite{2017paszkePyAD}. The backward mode is designed for an objective function with a large number of inputs and a small number of outputs~\cite{2018BaydinAD}, which is the case in this framework (thousands of inputs/scalar output).

AdamW~\cite{2017LoshchilovAdamW}, employs adaptive step size, momentum, and decoupled weight decay, which allows for improved convergence, especially for problems with sharp local minima. However, considering that the solved problem is deterministic and only one vector of variables is being optimized at a time, the optimizer does not provide a stochastic advantage. The proposed topology optimization benefits from the second moment estimate and from the decoupled weight decay, making the algorithm less prone to parameter choice. However, as a gradient descent algorithm, AdamW can diverge when optimizer hyperparameters are poorly selected. 

The bi-level optimization scheme is applied on top of the local optimization, which solves~\eqref{eq:opimization}. \gls{BO} chooses the optimal hyperparameters of the AdamW optimizer and the maximum values of the parameters of the projection methods (see Appendix~\ref{Appendix:proj}). Moreover, \gls{BO} sets the initial value of the vector of excitation coefficients~$\Vvec$~\eqref{eq:opimization}, thus defining the position of the delta-gap feeding~\cite{balanis2012advanced}. We will describe each part of the bi-level scheme separately.

\subsection{Auxiliary Edge Resistivity Optimization Algorithm}
\label{subsec:alg}
\begin{algorithm}[]
    \caption{Auxiliary edge resistivity TO}
    \label{alg:edgeres}
    \begin{algorithmic}[1]
        \REQUIRE \gls{MoM} operators, feeding~$\Vvec$, density filter, Q-factor lower bound $\Qlb$
        \STATE{\textbf{given:} Learning rate~$\alpha \in \mathbb{R}$, weight decay~$\lambda \in \mathbb{R}$, maximum thresholding factor~$\beta_\T{max} \in \mathbb{R}$, maximum regularization coefficients~$\gamma_\T{max}, \delta_\T{inc} \in \mathbb{R}$, maximum iterations~$i_\T{max}$ per $\beta$}
        \STATE{\textbf{initialize:} $\beta = 4$,~$\delta = 0$,~$\gamma = 0$ }
        \STATE{\textbf{initialize:} weights~$\V{w}\in\mathbb{R} ^{(N - d) \times 1}$, $\V{w} = 0.5$}, where~$d\in \mathbb{Z}$ is the number of fixed \gls{DOF}
        \REPEAT
            \STATE{$ i \leftarrow i + 1$}
            \STATE{$\ovart \leftarrow \sigma(\V{w})$} \COMMENT{$\sigma$ is the sigmoid function}
            \STATE{$\ovar \leftarrow \T{copy}(\ovart)$} \COMMENT{preserve values of fixed \gls{DOF}}
            \STATE{$\ovar \leftarrow f(\ovar)$} \COMMENT{apply projections}
            \STATE{$\projMatrix \leftarrow \ovar$} \COMMENT{Interpolation}
            \STATE{$\Ivec \leftarrow \projMatrix$} \COMMENT{Solve \gls{MoM}}
            \STATE{ $Q_\T{E},Q_\T{U} \leftarrow \Ivec$}
            \STATE{$\Loss \leftarrow Q_\T{E},Q_\T{U}, \ovar$}
            \IF{$i > i_\T{max}$}
                \IF{$\beta < \beta_\T{max}$}
                    \STATE{$ \beta \leftarrow 2\beta$}
                    \STATE{$ i \leftarrow 0$}
                    \STATE{$ \delta \leftarrow \delta + \delta_\T{inc}$}
                \ELSIF{$\gamma_\T{max} \neq 0$}
                    \STATE{$\gamma \leftarrow \gamma_\T{max}$}
                    \STATE{$ i \leftarrow 0$}
                \ELSE
                    \STATE{\textbf{break}}
                \ENDIF
            \ENDIF
            \STATE{Evaluate sensitivities using \gls{AD}}
            \STATE{AdamW step}
        \UNTIL{\textit{stopping criterion is met}}
        \RETURN{optimized weights $\V{w}$}
    \end{algorithmic}
\end{algorithm}

Algorithm~\ref{alg:edgeres} describes the auxiliary edge resistivity optimization routine. The algorithm is split into two main parts: pre-processing phase (steps $1$--$3$) and the main loop: steps ($4$--$25$.)

\subsubsection{The pre-processing}
In the pre-processing phase, we evaluate the \gls{MoM} operators, namely~$\Zmat, \Vvec, \XEmat, \XMmat$, and the density filter kernel~$\denstityKernel$. Furthermore, the Q-factor lower bound~$Q_\T{lb}^\T{TM}$ is computed.

The algorithm is initialized with vector of excitation coefficients~$\Vvec$, the optimizer learning rate~$\alpha \in \mathbb{R}$, the optimizer weight decay~$\lambda \in \mathbb{R}$ and the optimization parameters~$\beta_\T{max} \in \mathbb{R}$ representing the maximum value of the thresholding projection~\cite{2010projectionWang}, maximum number of steps~$i_\T{max}$ per~$\beta$, the maximum value of the regularization coefficient~$\gamma_\T{max} \in \mathbb{R}$ and the increment to the self-resonance coefficient $\delta_\T{inc} \in \mathbb{R}$. Lastly, the active regularization coefficients~$\beta$,~$\delta$,~$\gamma$ are set to their initial values.

\subsubsection{The main loop}
The main loop runs until~$\beta \geq \beta_\T{max} \land \gamma = \gamma_\T{max} \land i = i_\T{max}$. The terminating condition is achieved by periodically increasing the values of~$\beta$ and~$i$ until~$\beta \geq \beta_\T{max}$, after which~$\gamma$ is set to the maximum value and~$i$ is set to~$0$. Afterwards, the algorithm must optimize the loss function~$\Loss$ with the regularization term~\eqref{eq:regularization}, while~$\beta = \beta_\T{max}$ for~$i_\T{max}$ iterations. 

Unlike some optimizers, AdamW~\cite{2017LoshchilovAdamW} does not include constraint handling in its formulation. To enforce the box constraints on our design variable~$\ovar$ we utilize an unconstrained variable~$\V{w}\in\mathbb{R} ^{(N - d) \times 1}$, where~$d$ is the number of fixed \gls{DOF} and apply a sigmoid function~$\sigma$~\cite{1995HanSigmoid} which transforms~$\V{w}$ into~$\ovart$. The variables~$\ovart$ are then copied into~$\ovar$ preserving fixed \gls{DOF}, such as the delta-gap feeder. Afterward, the interpolation scheme is applied and the \gls{MoM} state equation is computed. Once the state equation is solved, the loss function~$\Loss$ is acquired. The stopping criterion is then checked and the regularization coefficients~$\delta$,~$\gamma$, and the projection parameter~$\beta$ are adjusted. Finally, the sensitivities are calculated using \gls{AD}, and a step using AdamW is evaluated. The algorithm returns the optimized weights~$\V{w}$. In order to get the design variable~$\ovar$ steps~$6$,~$7$,~$8$ must be applied to process~$\V{w}$.

\subsection{Bayesian Optimization}
\label{sec:bao}
\gls{BO} is a set of \gls{ML}-based optimization algorithms that maximize the value of the objective function~$f$~\cite{2018FrazierBO}. To efficiently apply \gls{BO},~$f$ should be \Quot{expensive to evaluate} and no derivatives should be observed~\cite{2018FrazierBO}. Generally, \gls{BO} is divided into two steps: building a Bayesian statistical model to model the objective
function~$f$ and using an acquisition function to decide where to sample next~\cite{1998JonesBO}. This paper uses a simple constrained \gls{BO} based on~\cite{2014FernandoBOPy}. 

Let us formulate the bi-level optimization problem for the optimizer hyperparameters and delta-gap feeding position as follows
\begin{equation}
    \label{eq:bo}
    \begin{aligned}
        \min_{\alpha, \lambda, i_{\T{max}}, p} \quad & \Lbo(\alpha, \lambda, i_{\T{max}}, p)  = \Loss(\ovar^{*}; \alpha, \lambda, i_{\T{max}}, p)\\
        \text{s.t.} \quad & \alpha \in [\alpha_{\min}, \alpha_{\max}], \\
                         & \lambda \in [\lambda_{\min}, \lambda_{\max}], \\
                         & i_{\T{max}} \in [i_{\min}, i_{\max}], \\
                         & \pn \in [0, p/(W/2 + L/2)] = [0, 1], \\
                         & \ovar^{*} = \arg\min_{\ovar} \Loss(\ovar; \alpha, \lambda, i_{\T{max}}, p_\T{N}),
    \end{aligned}    
\end{equation}
where~$\Lbo$ are the optimum results of the auxiliary edge resistivity optimization~\eqref{eq:opimization} conditioned on parameters~$\alpha$,~$\lambda$,~$i_{\T{max}}$, for optimal parameters~$\ovar^{*}$. 

\begin{figure}
\centerline{\includegraphics[width=\linewidth]{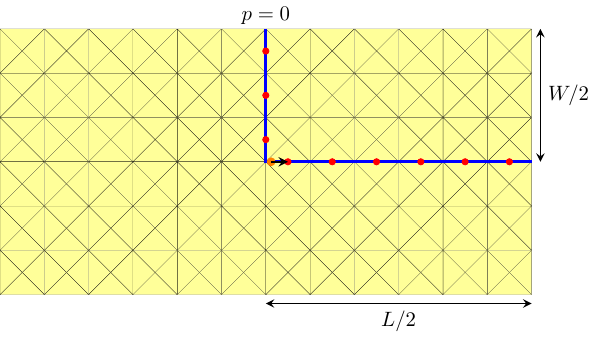}}
\caption{The parametrization of the delta-gap feeder. Normalized position parameter (suggested by \gls{BO})~$\pn \in [0,1]$ is re-scaled to $p \in [0, W/2 + L/2]$ and converted to a 2D Cartesian coordinate system, and the delta-gap feeder is found based on the second norm.}
\label{fig:pparametrization}
\end{figure}

The position of the delta-gap feeder is chosen based on a normalized parametrization of a line~$\pn$. Fig.~\ref{fig:pparametrization} shows the parametrization of the delta-gap feeder. The delta-gap feeder sits on the blue line. The position of the parametrized selected point (in orange) is converted to the 2D Cartesian coordinate system, and the second norm between the point and all edge centers sitting on the parametrized line (in red) is calculated. Let:
\begin{itemize}
    \item $\mathbf{p}_\mathcal{N} \in \mathbb{R}^2$ be the Cartesian coordinates of the parametrized position~$\pn$ on the line (orange point),
    \item $\mathbf{c}_n \in \mathbb{R}^2$ be the 2D coordinates of the center of edge $n$ (in red).
\end{itemize}
Then the index $n^*$ of the selected edge is given by
\begin{equation}
    n^* = \arg\min_{n = 1, \dots, N} \| \mathbf{p}_\mathcal{N} - \mathbf{c}_n \|_2
\end{equation}
The excitation vector $\mathbf{V}$ is defined as
\begin{equation}
    V_n = 
\begin{cases}
1, & \text{if } n = n^*, \\
0, & \text{otherwise}.
\end{cases}
\end{equation}

\section{Results}
\label{sec:results}
The proposed \gls{TO} algorithm is implemented in PyTorch~\cite{2017paszkePyAD} and the required \gls{MoM} matrices are computed in AToM~\cite{atom}. For the details of the implementation we refer the reader to the GitHub repository~\cite{2025BosakAERGitHub}. For the initial setting of the hyperparameters, we refer to Table~\ref{tab:rulesOfThumb} as it provides a good starting point to understand the algorithm. For optimum convergence, the readers are recommended to use \gls{BO} and setting~$\gamma_\T{max} = 1$, ~$\delta_\T{inc} = 0.2$,~$\beta_\T{max} = 64$ and changing~$r$ (radius of the density filter) with respect to the number of optimized variables. However, these settings are only recommended for minimizing the Q-factor and may fail in other optimization tasks.
\begin{table}[]
\caption{Recommended initial settings of the local algorithm.}
\begin{tabular}{cccccccc}
$\beta_\mathrm{i}/\beta_\mathrm{max}$ & $\alpha$ & $\lambda$ & $i_\mathrm{max}$ & $\gamma_\mathrm{max}$ & $\delta_\mathrm{inc}$ & $r$   & $R_\mathrm{o}/R_\mathrm{s}$ \\ \hline
$4/32$                                      & $0.7$    & $0.01$    & $75$             & $0$                   & $0$                   & $0.2a$ & $10^5/0.1$                 \\ \hline
\end{tabular}
\label{tab:rulesOfThumb}
\end{table}

\subsection{The Performance of the Local Algorithm}
The first set of examples compares the performance of the auxiliary edge resistivity algorithm with the material density-based~\gls{TO}~\cite{Q-TOPO}, binary \gls{TO}
based on a genetic algorithm~\cite{DebMultiOOusingEA} and \gls{TO} based on the memetic scheme~\cite{2021capeketalTSGAmemeticsPart1} that combines exact reanalysis~\cite[p. 128]{Ohsaki2016} and a
genetic algorithm. The results were obtained for two discretizations using~$1408$ and~$2454$ basis functions and aspect ratio $L/W = 5/3$~as used in~\cite[Section~V]{Q-TOPO}. The electrical size is set to $ka = 0.8$ in this example. A computer with~Intel~Xeon~Gold~6244~CPU ($16$~physical cores, $3.6$~GHz) with 384GB~RAM was utilized. The optimization hyperparameters were fixed to~$\delta_\T{inc} = 0$,~$\gamma_\T{max} = 0$, and~$\beta_\T{max} = 32$ to simulate the optimization function of~\cite[(15)]{Q-TOPO} for Q-factor minimization. In this work, the values of~$R_\T{o}$ and~$R_\T{s}$ are set based on the numerical investigations provided in~\cite{Q-TOPO} as~$R_\T{s} = 0.1$\,$\Omega$,~$R_\T{o} = 10^5$\,$\Omega$.

\begin{figure}[]
\centerline{\includegraphics[width=\linewidth]{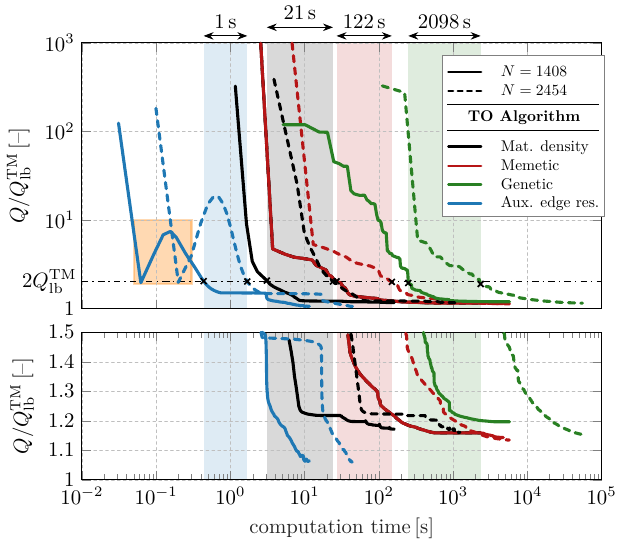}}
\caption{The performance of topology optimization algorithms for Q-factor
minimization for two discretizations with~$1408$ and~$2454$ basis functions. The Q-factor is
normalized to the \gls{TM} lower bound~$\Qlb$. Four implementations are shown: auxiliary edge resistivity topology optimization (blue curves), material density topology optimization~\cite{Q-TOPO} (black curves), a genetic algorithm~\cite{DebMultiOOusingEA} (green
curves), and a memetic algorithm~\cite{2021capeketalTSGAmemeticsPart1} (red curves).
Black crosses mark the computational time needed to reach~$Q = 2\Qlb$. The crosses demonstrate the scalability of the methods. The orange box indicates minor instability in the optimization caused by the nature of AdamW.}
\label{fig:timecomplexity}
\end{figure}

Fig.~\ref{fig:timecomplexity} depicts the comparison. The auxiliary edge resistivity outperforms all three algorithms in terms of convergence. Scalability is also compared using black marks and the black dash-dotted line, which indicates the time needed to cross the boundary~$Q = 2\Qlb$. The scalability of the auxiliary edge resistivity \gls{TO} is superior to the other \gls{TO} algorithms. The quantitative metrics are summarized in Table~\ref{tab:algComparison}.

Particular attention should be paid to the convergence behavior of the auxiliary edge resistivity \gls{TO}. Unlike the other methods, the auxiliary edge resistivity \gls{TO} is optimized using the adaptive \gls{GD} method AdamW~\cite{2017LoshchilovAdamW}. In general, \gls{GD} updates are influenced by historical gradient information retained by the optimizer. This influence is moderated in adaptive methods through the use of momentum and weight decay. However, divergence or instability can still occur (see the orange box in Fig.~\ref{fig:timecomplexity}), especially when interacting with projection operations or when the hyperparameters are not well tuned. This behavior is documented in~\cite{2017DiederikAdam} and~\cite{2017LoshchilovAdamW} and was previously analyzed in~\cite{2018ReddiAdamConvergence}. Instabilities in optimization are typically the largest for the second step when the momentum vectors are updated for the first time~\cite{2017LoshchilovAdamW}.

\begin{table}[]
\centering
\caption{Quantitative comparison of algorithms--including auxiliary edge resistivity, material density-based topology optimization~\cite{Q-TOPO}, memetic algorithm~\cite{2021capeketalTSGAmemeticsPart1}, and genetic algorithm~\cite{DebMultiOOusingEA}. This table summarizes the results presented in Fig.~\ref{fig:timecomplexity}.}
\label{tab:algComparison}
\resizebox{\columnwidth}{!}{%
\begin{tabular}{ccccc}
\multicolumn{1}{c}{}                                        & Aux. Edge Resistivity & Material TO & Memetic A. & Genetic A. \\ \hline
\multicolumn{5}{c}{$N = 1408$}                                                                                        \\ \hline
\multicolumn{1}{c}{$Q/Q_\mathrm{lb}^\mathrm{TM}$\,{[}--{]}} & $1.06$                & $1.17$     & $1.14$  & $1.2$   \\
\multicolumn{1}{c}{$t$\,{[}s{]}}                            & $12$                  & $161$      & $4748$  & $5762$  \\
\multicolumn{1}{c}{Iterations\,{[}--{]}}                           & $371$                 & $452$      & $68$    & $2385$  \\ \hline
\multicolumn{5}{c}{$N = 2454$}                                                                                        \\ \hline
\multicolumn{1}{c}{$Q/Q_\mathrm{lb}^\mathrm{TM}$\,{[}--{]}} & $1.06$                & $1.16$     & $1.14$  & $1.15$  \\
\multicolumn{1}{c}{$t$\,{[}s{]}}                            & $44$                  & $1233$     & $5732$  & $55208$ \\
\multicolumn{1}{c}{Iterations\,{[}--{]}}                           & $443$                 & $600$      & $132$   & $500$   \\ \hline
\end{tabular}%
}
\end{table}

\begin{figure}[]
\centerline{\includegraphics[width=\linewidth]{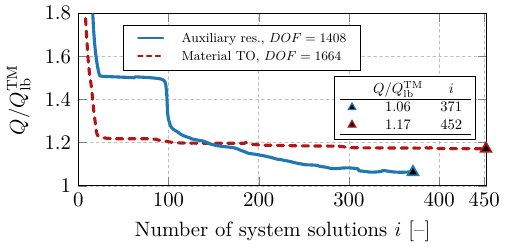}}
\caption{Convergence history of Q-factors normalized to the \gls{TM} Q-factor lower bound~\cite{2019CapekOA}. The result of auxiliary edge resistivities \gls{TO} is in blue and the result of the material density \gls{TO}~\cite{Q-TOPO} is in red. The auxiliary edge resistivities \gls{TO} utilizes~$1408$ \gls{DOF} and the material density \gls{TO} utilizes~$1664$. The triangular markers indicate when the algorithms reach the stopping criteria.}
\label{fig:convergencehistory}
\end{figure}

Fig.~\ref{fig:convergencehistory} shows the history of convergence of the material density-based \gls{TO}~\cite{Q-TOPO} (red curve) compared to the auxiliary edge resistivity \gls{TO} (blue curve). The auxiliary edge resistivities \gls{TO} utilizes~$1408$ \gls{DOF} and the material density \gls{TO} utilizes~$1664$. The triangular markers indicate when the edge resistivity algorithm reaches the stopping criteria. The radius of the density filter is the same for both methods as~$r = 0.15a$. The cost functions show that the density-based approach, utilizing \gls{MMA}~\cite{1987SvanbergMMA}, converges faster to the self-resonant design. However, the curves show that the edge resistivity approach utilizing~\cite{2017LoshchilovAdamW} reaches a better Q-factor ($11$\% improvement) than the density approach, with approximately~$200$ \gls{DOF} less than the material density approach and~$81$ system solutions less as indicated by the triangular markers. It should also be pointed out that the Algorithm~\ref{alg:edgeres} does not involve adaptive updates of the thresholding scheme~\eqref{eq:cprojection} based on updates of the optimized variable. The thresholding parameter is updated after~$i_\T{max} = 62$ iterations. The blue curve also shows that the convergence is stopped between approximately~$25$th and~$95$th iterations.

\begin{figure}[]
\centerline{\includegraphics[width=\linewidth]{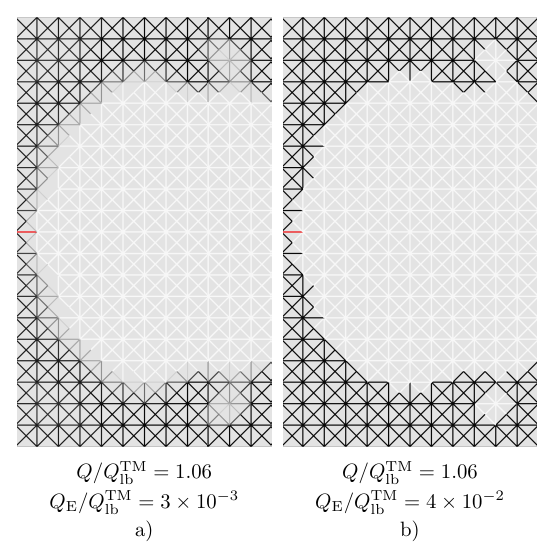}}
\caption{The results of auxiliary edge resistivity topology optimization for Q-factor
minimization for~$1408$ basis functions. a) shows the grayscale design, b)~shows the thresholded design. Electrical size is $ka = 0.8$. The Q-factor~$Q$ and tuning Q-factor~$\Qe$ are normalized to the \gls{TM} lower bound~$\Qlb$. Time needed to reach the convergence is~$12$\,s.}
\label{fig:convergences}
\end{figure}

Fig.~\ref{fig:convergences} shows the grayscale and thresholded convergence depicted by the blue curve in Fig.~\ref{fig:convergencehistory}. The Q-factors normalized by the \gls{TM} lower bound~$\Qlb$ are~$Q/\Qlb = 1.06$, $\Qe/\Qlb = 3\times 10^{-3}$ for the grayscale design and~$Q/\Qlb = 1.06$, $\Qe/\Qlb = 4\times 10^{-2}$ for the threshold design. Material density-based \gls{TO}~\cite{Q-TOPO} achieved the value of~$Q/\Qlb = 1.06$ in~$8.1\times10^{3}$\,s with~$5696$ basis functions. The auxiliary edge resistivity achieved~$Q/\Qlb = 1.06$ in~$12$\,s.

\subsection{Bi-level Optimization Results}
\begin{figure}
\centerline{\includegraphics[width=\linewidth]{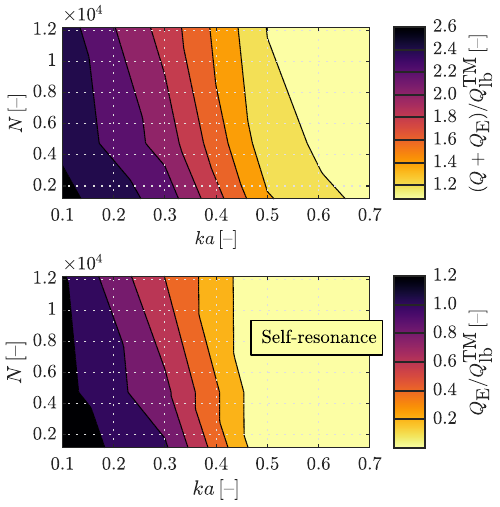}}
\caption{Self-resonant Q-factor optimization for electrically small antennas for~$ka \in [0.1, 0.7]$ and size ratio~$L/W = 2$. The top pane: Q-factor with the untuned~$Q$ and tuning~$\Qe$ Q-factors normalized by the TM lower bound. The bottom pane: tuning Q-factor~$\Qe$ normalized by the TM lower bound. }
\label{fig:restot}
\end{figure}

We further investigate the rectangular design region with dimension ratio~$L/W = 2$ and electrical sizes~$ka = \{0.1, 0.3, 0.5, 0.7\}$. The setting of the fixed hyperparameters was~$\beta_\T{max} = 64$, $\delta_\T{inc} = 0.3$, $\gamma_\T{max} = 1$. Fig.~\ref{fig:restot} depicts the total acquired results of the bi-level optimization. The upper panel shows the Q-factor normalized by the \gls{TM} lower bound. When the electrical size~$ka$ decreases, the optimization requires a large number of \gls{DOF} to reach the \gls{TM} lower-bound and achieve self-resonance. The results presented are comparable with the meanderline antennas from~\cite{2019CapekOA} with electrical size~$ka = 0.5$ and higher. For antennas with~$ka < 0.5$, a larger number of \gls{DOF} than~$1.2\times10^4$ is required. For antennas with~$\Qe \neq 0$ (non-self-resonating antennas), external tuning circuits are necessary.

\begin{figure*}
\centerline{\includegraphics{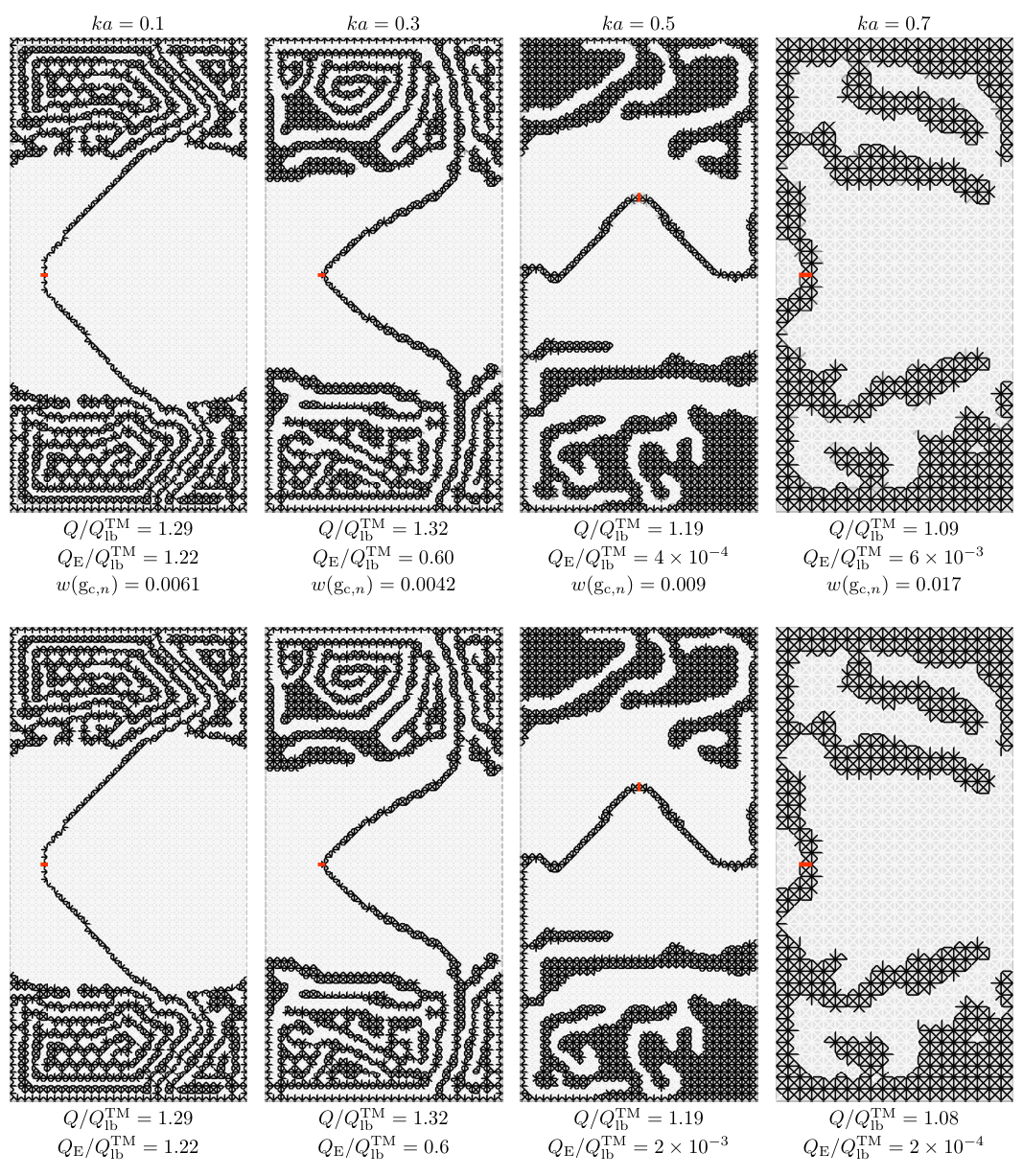}}
\caption{A matrix of the best achieved results for electrical sizes~$ka \in \{0.1,0.3,0.5,0.7\}$. The top row shows the non-thresholded results with the binarization score $\w$ (regularization function) and the respective Q-factors $\Qe$ and $Q$ normalized by the TM lower bound~$\Qlb$. The bottom row shows the thresholded ($\w = 0$) topologies and the respective Q-factors.}
\label{fig:restotmatrix}
\end{figure*}

Fig.~\ref{fig:restotmatrix} depicts the best achieved results for electrical sizes~$ka \in \{0.1,0.3,0.5,0.7\}$. All antennas were designed for frequency~$f_0 = 1$\,GHz. The upper row shows the non-thresholded results with the binarization score $\w$ \eqref{eq:regularization} and the respective Q-factors~$\Qe$~\eqref{eq:qe} and~$Q$~\eqref{eq:qu} normalized by the \gls{TM} lower bound~$\Qlb$. The bottom row shows the thresholded~($\w = 0$) topologies and the respective Q-factors. The red line indicates the position of the delta-gap feeding. All results are using the dimension ratio $L/W = 2$ and except for the antenna with electrical size~$ka = 0.7$, approximately $1.2\times10^4$ \gls{DOF} were utilized. The antenna shape with electrical size~$ka = 0.7$ uses roughly~$0.5\times10^4$ \gls{DOF}. 

Compared to meanderline antennas~\cite{2019CapekOA}, in terms of~$Q$, the antenna with electrical size~$ka = 0.7$ outperforms the meanderline antenna by roughly~$10$\%. The antenna at~$ka = 0.5$ is only slightly worse (roughly~$3$\%), and the remaining antennas at electrical sizes~$ka \in \{0.1, 0.3\}$ remain beaten by the meanderline antennas. However, as documented in Fig.~\ref{fig:restot}, this is only a problem of insufficient amount of \gls{DOF} and finer discretization than~$N = 1.2\times10^4$ should be used for antenna structures with~$ka \leq 0.5$. 

Self-resonance according to the tuning Q-factor~$\Qe \leq 0.1$ was achieved for electrical sizes~$ka \geq 0.5$, which can also be observed in Fig.~\ref{fig:restot}. Due to strong regularization~\eqref{eq:regularization} the \gls{TO} produces near-binary topologies and Q-factor changes minimally from the grayscale representation. The final values of~$\w$ and the visual representation demonstrate the difference between the grayscale and thresholded topologies. Fig.~\ref{fig:restotmatrix} shows that as the value of~$w$ increases the differences in the Q-factors between the grayscale and the thresholded ones increase.

\begin{table*}[]
\centering
\caption{Summary of achieved results for electrical sizes~$ka \in \{0.1,0.3,0.5,0.7\}$. The top sub-table shows the initial conditions and results for all electrical sizes. The bottom sub-table summarizes the achieved results by using Bayesian optimization. The respective columns are the number of degrees of freedom, the learning rate, the weight decay, the maximum number of steps per projection steepness, the port position, the computing time, the loss, the steps of Bayesian optimization, and the improvement of loss.}
\begin{tabular}{cccccccccc}
\multicolumn{9}{c}{Initial design} \\ \hline
\multicolumn{1}{c}{$ka$\,[--]} & \multicolumn{1}{c}{$N$\,[--]} &\multicolumn{1}{c}{$\alpha$\,[--]} & \multicolumn{1}{c}{$\lambda$\,[--]} & \multicolumn{1}{c}{$i_\T{max}$\,[--]} & \multicolumn{1}{c}{$\pn$\,[--]} & \multicolumn{1}{c}{Computing time\,[s]} & \multicolumn{1}{c}{$\Loss$\,[--]} & \multicolumn{1}{c}{BO steps\,[--]} & \multicolumn{1}{c}{$\Delta\Loss$\,[\%]} \\ \hline
$0.1$ & $12192$ & $5\times10^{-2}$ & $2\times10^{-3}$ & $2220$ & $0$ & $2039$ & $2.78$ & $1$ & $0$\\
$0.3$ & $12192$ & $5\times10^{-2}$ & $2\times10^{-3}$ & $2220$ & $0$ & $2039$ & $2.10$ & $1$ & $0$\\
$0.5$ & $12192$ & $5\times10^{-2}$ & $2\times10^{-3}$ & $2220$ & $0$ & $2039$ & $1.31$ & $1$ & $0$\\
$0.7$ & $4740$ & $5\times10^{-2}$ & $2\times10^{-3}$ & $2220$ & $0$ & $768$ & $1.13$ & $1$ & $0$\\ \hline
\multicolumn{9}{c}{Final design}\\ \hline
$0.1$ & $12192$ & $3\times10^{-2}$ & $4\times10^{-3}$ & $2232$ & $0.09$ & $23858$ & $2.75$ & $11$ & $3.1$\\
$0.3$ & $12192$ & $1.3\times10^{-2}$ & $2\times10^{-3}$ & $1592$ & $0.15$ & $11370$ & $2.04$ & $6$ & $5.4$\\
$0.5$ & $12192$ & $3\times10^{-2}$ & $4\times10^{-3}$ & $1947$ & $0.55$ & $31267$ & $1.20$ & $16$ & $11.6$\\
$0.7$ & $4740$ & $1\times10^{-2}$ & $1\times10^{-3}$ & $2044$ & $0.08$ & $9012$ & $1.08$ & $13$ & $5.3$\\ \hline
\end{tabular}
\label{tab:bo-table}
\end{table*}

Table~\ref{tab:bo-table} provides further quantitative results. All results were obtained using the NVIDIA A100-PCIE-40GB graphics card. The maximum improvement was achieved for the electrical size~$ka = 0.5$ with~a~$11.6$\% difference from the initial topology. However, the computing time in this case was also the highest, reaching more than~$8.6$~hours in~$16$ steps of \gls{BO}. As documented by the port position values~$\pn$ and Fig.~\ref{fig:restotmatrix}, the optimal position of the delta-gap feeding according to \gls{BO} does not lie in the center of the longer edge~$\pn = 0$.

\subsection{Pre-Manufacturing Models}
\label{subsec:cst}
The output of the auxiliary edge resistivity \gls{TO} is an \gls{EFIE} \gls{MoM} compatible thresholded vector~$\ovar$, see Algorithm~\ref{alg:edgeres}. The complete removal of a piece of material (a triangle in the \gls{RWG} mesh, see Fig.~\ref{eq:momrelaxed}) requires the removal of all edges~\cite{CapeketalShapeSynthesisBasedOnTopologySensitivity} (see Fig.~\ref{fig:rwgrelaxation}) assigned to that piece of material. However, as long as an edge is present, there must be two pieces of material connected to this edge~\cite{CapeketalShapeSynthesisBasedOnTopologySensitivity}. Consequently, the auxiliary edge resistivity \gls{TO} can generate manufacturing-unfriendly artifacts. The most salient are point connections and infinitely thin slots~\cite{2025FriendlyShapedNeuman}, which have to be added if two adjacent triangles have their common \gls{RWG} edge removed. The use of density filters suppresses checkerboard patterns~\cite{2001BrunselasticStructures}, but leaves intermediate material densities between material and void~\cite{Haber1996}\cite{Q-TOPO}, leading to intermediate edge resistivities, thus slots. Due to the point connections and infinitely thin slots, the output of the \gls{TO} requires a treatment and subsequent verification of the performance.

The antenna designs generated by the auxiliary edge resistivity \gls{TO} for electrical sizes~$ka \in \{0.5, 0.7\}$ were further investigated in terms of Q-factor. We compare two Q-factors -- the self-resonant Q-factor~$Q$~\eqref{eq:q}, and~$\Qb$, which is determined from \gls{FBW} and \gls{RL} of an antenna~\cite{YaghjianBestImpedanceBandwidthAndQOfAntennas}. The Q-factor~$\Qb$ is evaluated as~\cite{YaghjianBestImpedanceBandwidthAndQOfAntennas}
\begin{equation}
    \Qb = \frac{2 \, \T{RL}_\T{level}}{\T{FBW}\sqrt{1-(\T{RL}_\T{level})^2}},
\end{equation}
where~$\T{RL}_\T{level}$ is set to~$-3$\,dB, and assuming an electrically small, single-port, and matched antenna,~$Q \approx \Qb$,~\cite{YaghjianBestImpedanceBandwidthAndQOfAntennas}. Considering that \gls{RL} is not an optimized quantity, the real part of the input impedance~$\Zin$ can differ from the standard characteristic impedance of transmission lines such as~$50$\,$\Omega$. If~$\Xin \neq 0$ ($\Qe \neq 0$) then an additional series tuning inductor or a capacitor is needed to tune the antenna to a resonance~\cite{YaghjianBestImpedanceBandwidthAndQOfAntennas}. \gls{RL} is evaluated as
\begin{equation}
    \T{RL}(f) = \left| \frac{\Zin(f) + \J X_\T{a} - R_0}{\Zin(f) + \J X_\T{a} + R_0}\right |^2,
\end{equation}
where we substitute~$R_0 = \RE\{\Zin(f_0)\}$ as the value of the characteristic impedance of the transmission line, and~$X_\T{a}$ is the reactance of the series tuning element substituted as~$X_\T{a} = -\IM\{\Zin(f_0)\}$.

The further assessed results show a comparison of the \gls{RL} of the antenna structure without any ($X_\T{a} = 0$), with a tuning circuit ($X_\T{a} = -\IM\{\Zin(f_0)\})$, and a verification in CST Microwave Studio~\cite{cst} without a tuning circuit ($X_\T{a} = 0$). The designs in CST were evaluated using the Time Domain Solver. The designs for~$ka = \{0.5, 0.7\}$ are assessed in the following subsections. 

For the radiation characteristics of the antennas, refer to Appendix~\ref{Appendix:patterns}.

\subsection{Optimal Antenna Sample at $ka = 0.5$}
The operational frequency was set to~$f_0 = 1$\,GHz. The input impedance simulated by AToM \gls{MoM}~\cite{atom} in~$f_0$ was evaluated as~$\Zin = 7.519 - \J 0.889$\,$\Omega$. An additional inductive element had to be added to compensate for the capacitive character of the reactance.

\begin{figure}[]
\centerline{\includegraphics[width=\linewidth]{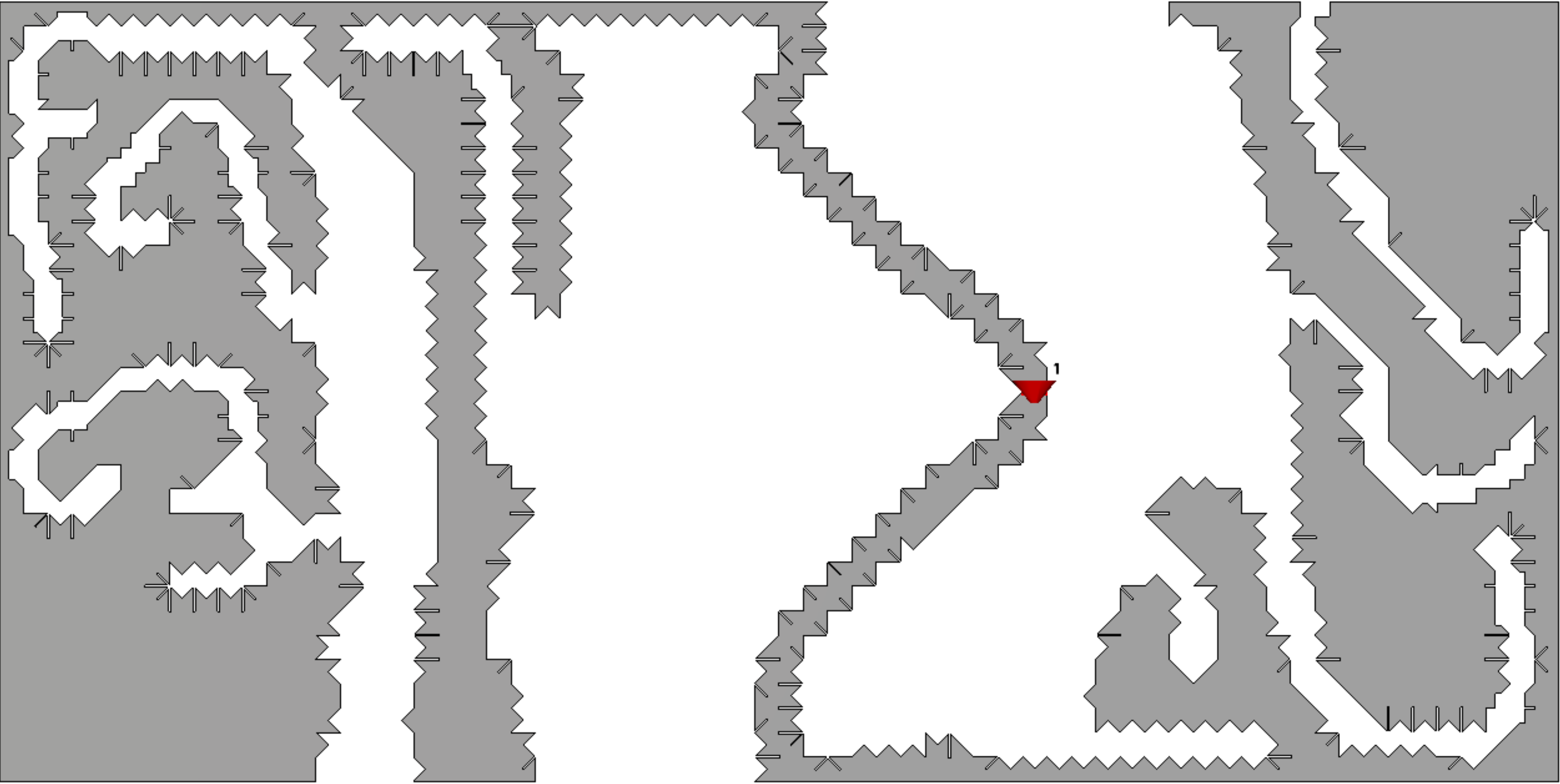}}
\caption{The design for~$ka = 0.5$ and~$f_0 = 1$\,GHz converted into CST Microwave Studio.}
\label{fig:aer3D}
\end{figure}

Fig.~\ref{fig:aer3D} shows the converted design in CST Microwave Studio. The node connections were removed, and infinitely thin slots were replaced with slots of finite width $0.07$\,mm.

\begin{figure}[]
\centerline{\includegraphics[width=\linewidth]{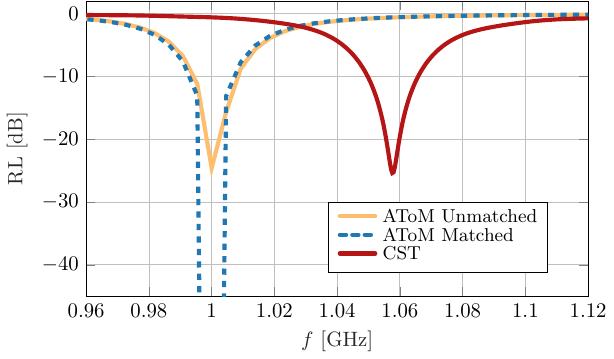}}
\caption{Return loss comparison for an antenna at~$ka = 0.5$, shown in Fig.~\ref{fig:aer3D}, across the results obtained in AToM without a tuning circuit (orange curve), with a tuning circuit (blue), and CST (red curve).}
\label{fig:RL}
\end{figure}

Fig.~\ref{fig:RL} shows the comparison of AToM, AToM result matched, and CST. The matching of the input impedance significantly improves the performance of \gls{RL} at the design frequency of~$1$\,GHz. However, \gls{FBW} remains almost constant, and no frequency shift occurs. The CST model shows an approximately~$60$\,MHz frequency shift of the resonance, which means that the electrical size~$ka$ changed to~$ka = 0.531$.  

Table~\ref{tab:ka05results} summarizes the results achieved and depicted in Fig.~\ref{fig:RL}. As expected, the tuned antenna matches the performance of the untuned antenna in terms of fractional bandwidth and~$\Qb$. The converted antenna shows slightly larger \gls{FBW} and lower~$\Qb$ with approximately~$0.37$\% improvement and the expected frequency shift of~$60$\,MHz. 

\begin{table}[]
\centering
\caption{The results of a design at~$ka = 0.5$ and~$f_0 = 1$\,GHz, showed in Fig.~\ref{fig:aer3D}, compared across AToM, AToM result matched with inductor and CST Microwave Studio.}
\begin{tabular}{ccccc}
Design       & $f_0$\,[GHz] & FBW\,[--] & $Q$\,[--] & $\Qb$\,[--] \\ \hline
AToM         &  $1$     &  $0.0411$   & $49.175$ & $48.827$  \\
AToM Matched &  $1$     &  $0.0412$   & $49.234$ & $48.612$ \\
CST          &  $1.06$  &  $0.0449$   &  --      & $44.635$ \\ \hline
\end{tabular}
\label{tab:ka05results}
\end{table}

\subsection{Optimal Antenna Sample at $ka = 0.7$}
\begin{figure}[]
\centerline{\includegraphics[width=\columnwidth]{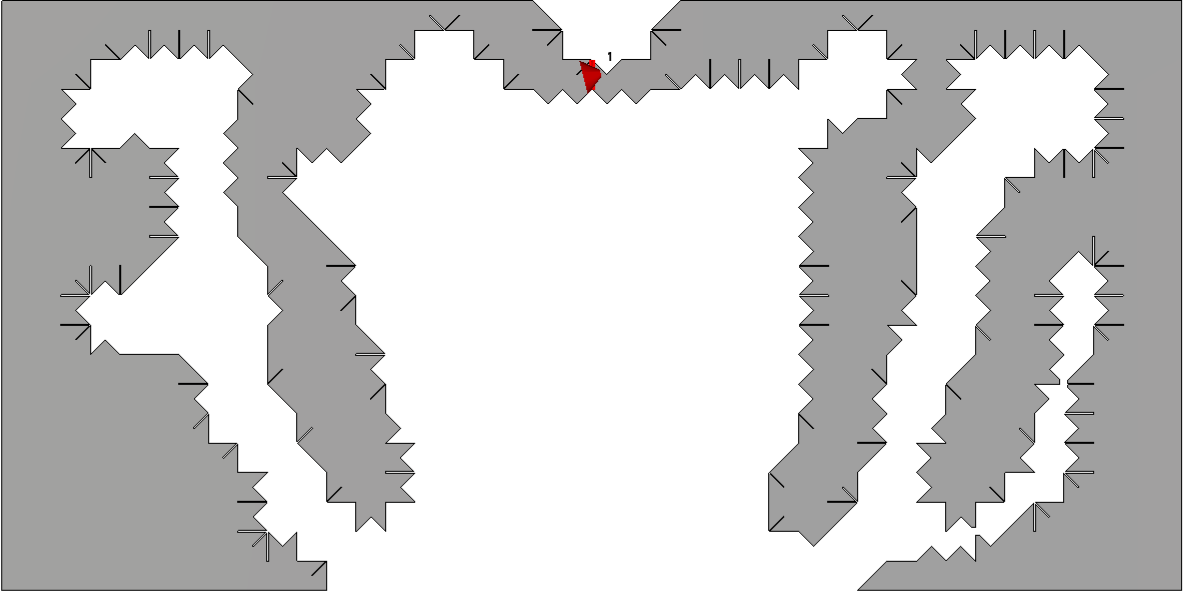}}
\caption{Design for $ka = 0.7$ and $f_0 = 1$~GHz converted into CST Microwave Studio.}
\label{fig:aer3D07}
\end{figure}

The converted antenna is depicted in Fig.~\ref{fig:aer3D07}. The input impedance was evaluated as~$\Zin = 18.275 - \J1.978$\,$\Omega$, and an inductive element was added to compensate for the reactance. The width of the finite slots is $0.08$\,mm.

\begin{figure}[]
\centerline{\includegraphics[width=\linewidth]{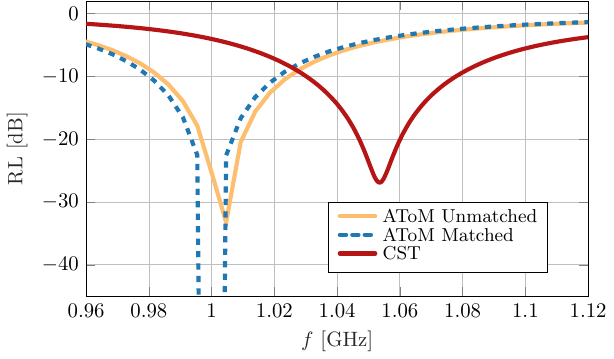}}
\caption{Return loss comparison for an antenna at~$ka = 0.7$, showed in Fig.~\ref{fig:aer3D07}, across the results obtained in AToM without a tuning circuit (orange curve), with a tuning circuit (blue), and CST (red curve).}
\label{fig:RL07}
\end{figure}

Fig.~\ref{fig:RL07} shows the comparison across AToM, AToM result matched, and CST. The matching of the input impedance significantly improves the performance of \gls{RL} at the design frequency of $1$\,GHz. In this case, \gls{FBW} remains almost constant, but there is a slight frequency shift of approximately $5$\,MHz. The CST model shows an approximately $55$\,MHz frequency shift of the resonance, which means that the electrical size~$ka$ shifted to~$ka = 0.738$.  

Table~\ref{tab:ka07results} summarizes the results achieved and depicted in Fig.~\ref{fig:RL07}. As expected, the tuned antenna matches the performance of the untuned antenna in terms of fractional bandwidth and~$\Qb$. The converted antenna shows slightly larger \gls{FBW} and lower~$\Qb$ with approximately~$1.4$\% improvement and the expected frequency shift of~$55$\,MHz. 

\begin{table}[]
\centering
\caption{The results of a design at $ka = 0.5$ and $f_0 = 1$\,GHz compared across AToM, AToM result matched with inductor and CST Microwave Studio.}
\begin{tabular}{ccccc}
Design       & $f_0$\,[GHz] & FBW\,[--]  & $Q$\,[--]  & $\Qb$\,[--] \\ \hline
AToM         &  $1$     &  $0.1242$     & $16.211$  & $16.147$  \\
AToM Matched &  $1.005$     &  $0.1238$     & $16.265$  & $16.190$ \\
CST          &  $1.055$  &  $0.1373$    & --        & $14.601$ \\ \hline
\end{tabular}
\label{tab:ka07results}
\end{table}

\section{Conclusions}
\label{sec:conclusions}
The article presented a novel gradient-based topology optimization approach within the method-of-moments paradigm. The method utilizes auxiliary variables called edge resistivities to optimize shape topologies using adaptive gradient descent and automatic differentiation. Furthermore, the projection operators known in topology optimization are reformulated for auxiliary edge resistivities, and a new regularization term is added to the optimized function to ensure a binary result of the optimization. The bi-level optimization scheme based on Bayesian optimization is applied for finding the optimal position of the delta-gap feeder and optimal optimizer hyperparameters. 
The performance of the method was verified on Q-factor minimization, a well-understood problem, with tightly defined lower bounds for electrically small antennas. 

The paper presented numerical investigations in rectangular optimization regions. The new local method outperforms current state-of-the-art methods in terms of convergence. However, due to the nature of gradient descent, careful tuning of the optimizer hyperparameters is required. The global algorithm further improves the robustness of the optimization and improves the convergence quality, but increases the optimization time. We compared the solutions of our method to meanderline antennas. The proposed topology optimization exceeds the meanderline antennas for electrical sizes larger than~$0.5$. For electrical sizes lower than~$0.5$, the meanderline antennas perform better. However, a finer discretization improves the Q-factor as proved by the numerical results. The proposed method enables optimizing an extreme number of degrees of freedom (tens of thousands), unlike genetic algorithms or memetic schemes. Therefore, a finer discretization should pose a minimum challenge in terms of computing time. 

The converted antennas show a remaining problem in the Q-factor optimization -- frequency shifts that occur due to slots and node connections generated in the RWG mesh. The two converted designs exhibit a~$60$\,MHz frequency shift in the resonance. 

Future research involving the proposed approach should revolve around multicriteria optimization involving the matching of the antenna, the antenna realized gain, and the structural parameters such as the number of slots and the number of node connections in the design, preventing drops in the performance when the antenna structure is converted to a pre-manufacturing model. The method can be easily extended to volumetric problems, and further modifications can be done, allowing an optimization of multi-port antennas. The applied automatic differentiation scheme allows for further exploration of different penalization functions and varying optimization algorithms, increasing the number of possibilities for improved convergence. The method can also serve as a generator of an initial design for further optimization tactics.

\appendices
\section{Projection Methods}
\label{Appendix:proj}
The first in the composition of projection functions~$f$ is the penalization function~\cite{topoOpt}
\begin{equation}
    \label{eq:penalization}
    f_1(\ovarn) = \frac{\ovarn}{1 + p(1 - \ovarn)},
\end{equation}
$p$ is the penalty parameter fixed throughout the optimization to~$p = 1$. The function forces the optimization algorithm to converge to near-binary values.

To improve the quality of convergence and the robustness of the optimization routine, density filters are introduced~\cite{2001BourdinDensityFilters}\cite{2001BrunselasticStructures}. The density filter is defined per element~$\ovarn$ as~\cite{2001BourdinDensityFilters} 
\begin{equation}
    \label{eq:filter}
    f_2(\ovarn) = \sum\limits_{m} \denstityKerneln \T{g}_{\T{c},m},
\end{equation}
where $\denstityKerneln\in\mathbb{R}$ is the filter kernel defined element-wise
\begin{equation}
    \label{eq:fkernel}
    \denstityKerneln = 
    \begin{cases}
    \dfrac{r - \norm{\V{c}_n - \V{c}_m}}{\sum\limits_{\V{c}_k\in \mathcal{N}_r(\V{c}_m)} (r - \norm{\V{c}_n - \V{c}_k})}, \quad & \V{c}_n \in \mathcal{N}_r(\V{c}_m), \\[5ex]
    
    0, \quad & \V{c}_n \notin \mathcal{N}_r(\V{c}_m),
    \end{cases}
\end{equation}
where~$\V{c}_m$ stands for a triangle edge center. The neighborhood domain $\mathcal{N}_r$ is a sphere of radius~$r$ that encircles the point~$\V{c}_m$. The definition of the domain is 
\begin{equation}
    \mathcal{N}_r(\mathbf{c}_m) = \left\{ \mathbf{c}_n \in \mathbb{R}^d \;:\; \norm{ \mathbf{c}_n - \mathbf{c}_m} \leq r \right\} .
\end{equation}
This allows control of the minimum feature size in the optimization. Radius~$r$ becomes a hyperparameter of optimization, but is often related to the manufacturing process for the final result~\cite{topoOpt}. The density filter introduces intermediate edge resistivities. 

To push the optimization towards a binary result, a thresholding projection scheme is introduced~\cite[(9)]{2010projectionWang},
\begin{equation}
    \label{eq:cprojection}
    f_3(\ovarn) = \frac{\tanh(\beta\eta) + \tanh(\beta(\ovarn - \eta)}{\tanh(\beta\eta) + \tanh(\beta(1-\eta))},
\end{equation}
where~$\beta$ is the steepness parameter and~$\eta$ is the projection level parameter. The projection level parameter is fixed as~$\eta = 0.5$ for all subsequent tasks. The parameter~$\beta$ is gradually increased throughout optimization to force the optimization to converge to near-binary results. The actual binary result is achieved by applying the Heaviside function instead, which is applied at the end of the optimization to verify the results. However, such a projection is impossible through optimization because the Heaviside function has no finite derivative. 

\section{Q-factor Evaluation}
\label{Appendix:q}
For the given current expansion coefficients~$\Ivec$,~$\We$ and~$\Wm$ represent stored electric and magnetic energies which can be written as~\cite{Vandenbosch2010}, \cite{GustafssonTayliEhrenborgEtAlAntennaCurrentOptimizationUsingMatlabAndCVX}
\begin{equation}
\begin{aligned}
\label{eq:wmats}
\We &\approx \frac{1}{4\omega}\Ivec^{\herm}\XEmat\Ivec, \\
\Wm &\approx \frac{1}{4\omega}\Ivec^{\herm}\XMmat\Ivec.
\end{aligned}
\end{equation}
Operator~$\herm$ stands for the Hermitian transpose, and~$\omega$ represents the angular frequency. The matrices~$\XEmat$ and~$\XMmat$ are the electric and magnetic reactance matrices. The matrices $\XEmat$ and $\XMmat$ are obtained as~\cite{GustafssonTayliEhrenborgEtAlAntennaCurrentOptimizationUsingMatlabAndCVX}
\begin{equation}
\begin{aligned}
\label{eq:xmats}
\XMmat &= \frac{1}{2}(\Wmat + \Xmat), \\
\XEmat &= \frac{1}{2}(\Wmat - \Xmat),
\end{aligned}
\end{equation}
where~$\Wmat$ represents the stored energy matrix which is for~$ka < 1$ defined as~\cite{Harrington1972}
\begin{equation}
    \Wmat = \IM\left\{\omega\frac{\partial\Zmat}{\partial\omega}\right\}.
\end{equation}
The imaginary part of the \gls{EFIE} impedance matrix~$\Zmat$ can be rewritten as~$\Xmat = \XMmat - \XEmat$. 

The term~$\Prad$ stands for the radiated power, which is defined as
\begin{equation}
    \label{eq:prad}
    \Prad \approx \frac{1}{2}\Ivec^{\herm}\Rmat\Ivec,
\end{equation}
where~$\Rmat$ is the real part of the \gls{EFIE} vacuum impadance matrix~$\Zmat$.

\section{Pre-Manufacturing Models: Radiation Patterns}
\label{Appendix:patterns}
\begin{figure}[]
\centering
\includegraphics[width=0.6\linewidth]{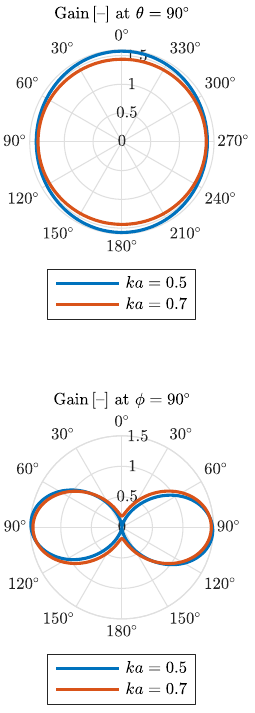}
\caption{Gain patterns for optimal antennas at $ka = 0.5$,~$ka = 0.7$ placed in the~$x$--$z$ plane with the longer side parallel to the~$z$~axis, and centered with respect to the coordinate origin.}
\label{fig:RadiationPatterns}
\end{figure}

Fig.~\ref{fig:RadiationPatterns} presents the radiation patterns of the optimized antennas. The antennas are placed in the~$x$--$z$ plane with the longer side parallel to the~$z$~axis, and centered with respect to the coordinate origin. The antennas predominantly radiate in $\hat{\V{\theta}}$ polarization. In general, the radiation characteristics closely resemble those of a dipole antenna~\cite{BalanisModernAntennaHandbook}. The far-field patterns were acquired using CST.

\begin{IEEEbiography}[{\includegraphics[width=1in,height=1.25in,clip,keepaspectratio]{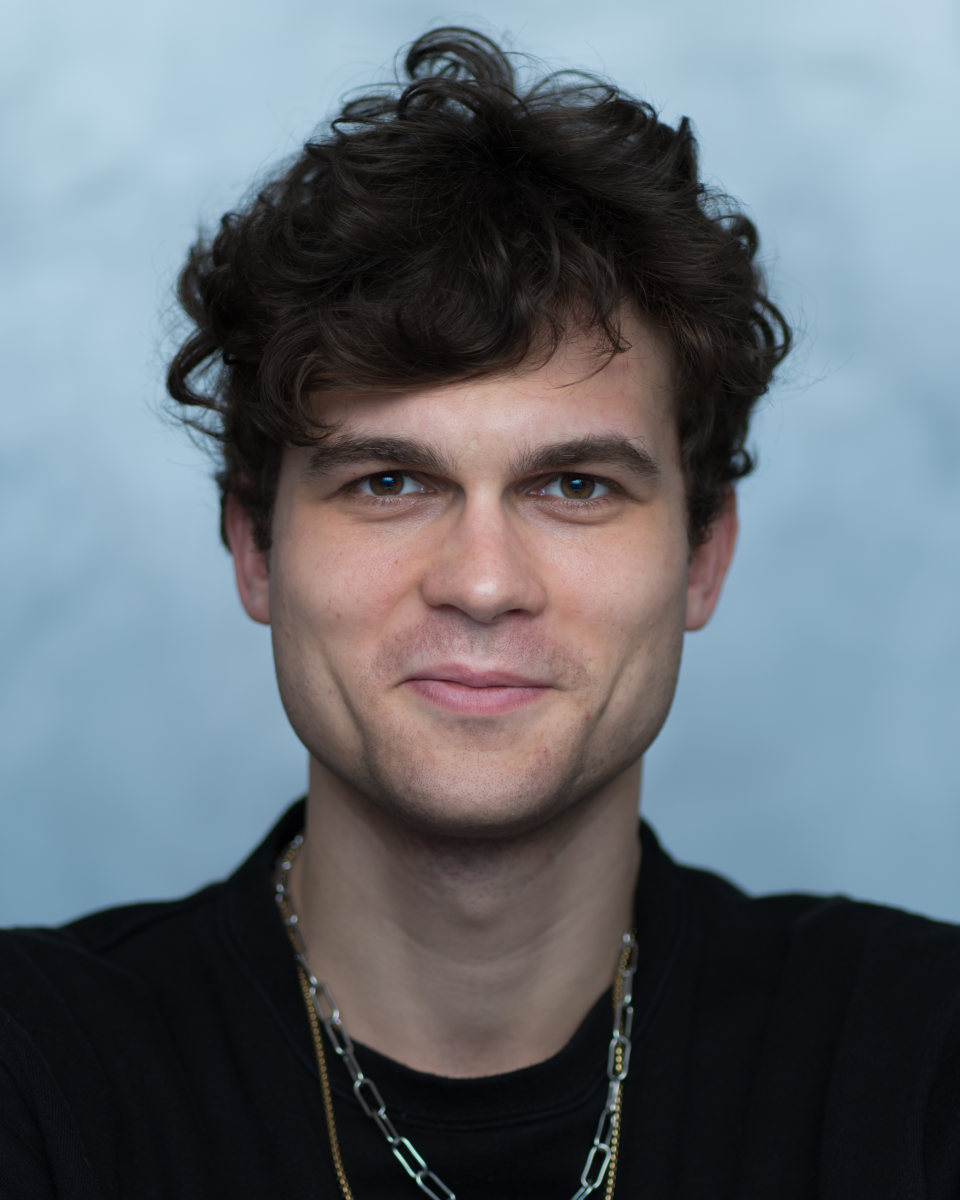}}]{Stepan Bosak } received his first M.Sc. degree in electronic and computer engineering from National Taiwan University of Science and Technology, Taipei, ROC, in 2022 and his second M.Sc. degree in radio engineering from the Czech Technical University in Prague, Prague, Czech Republic, in 2023. He is currently pursuing a Ph.D. in the area of Machine-Learning-Assisted Inverse Design in Electromagnetism. 
\end{IEEEbiography}

\begin{IEEEbiography}[{\includegraphics[width=1in,height=1.25in,clip,keepaspectratio]{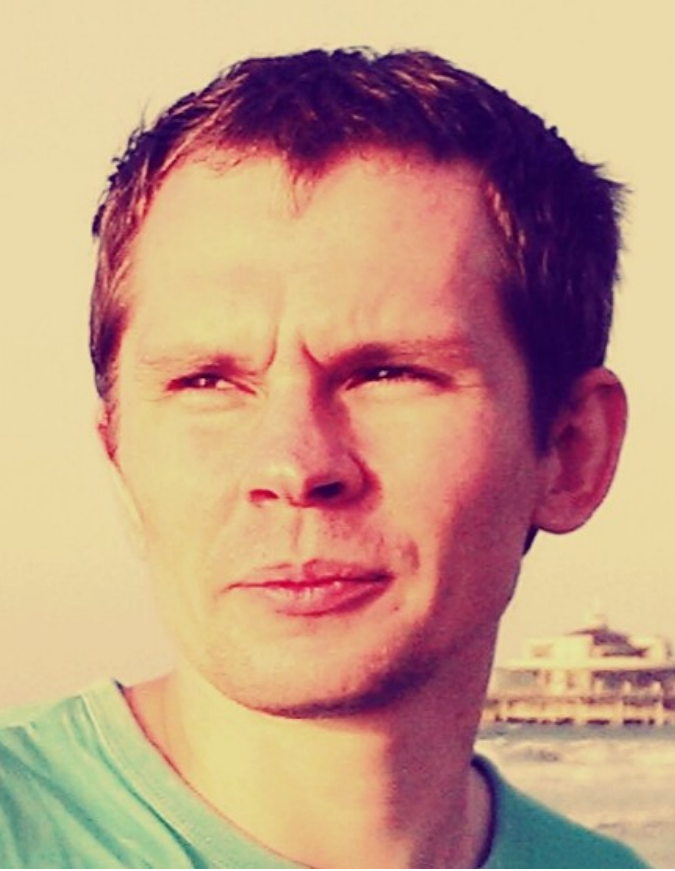}}]{Miloslav Capek }
(M'14, SM'17) received the M.Sc. degree in Electrical Engineering 2009, the Ph.D. degree in 2014, and was appointed a Full Professor in 2023, all from the Czech Technical University in Prague, Czech Republic.
	
He leads the development of the AToM (Antenna Toolbox for Matlab) package. His research interests include electromagnetic theory, electrically small antennas, antenna design, numerical techniques, and optimization. He authored or co-authored over 165~journal and conference papers.

Dr. Capek is the Associate Editor of IET Microwaves, Antennas \& Propagation. He was a regional delegate of EurAAP between 2015 and 2020 and an associate editor of Radioengineering between 2015 and 2018. He received the IEEE Antennas and Propagation Edward E. Altshuler Prize Paper Award~2022 and ESoA (European School of Antennas) Best Teacher Award in~2023.
\end{IEEEbiography}

\begin{IEEEbiography}[{\includegraphics[width=1in,height=1.25in,clip,keepaspectratio]{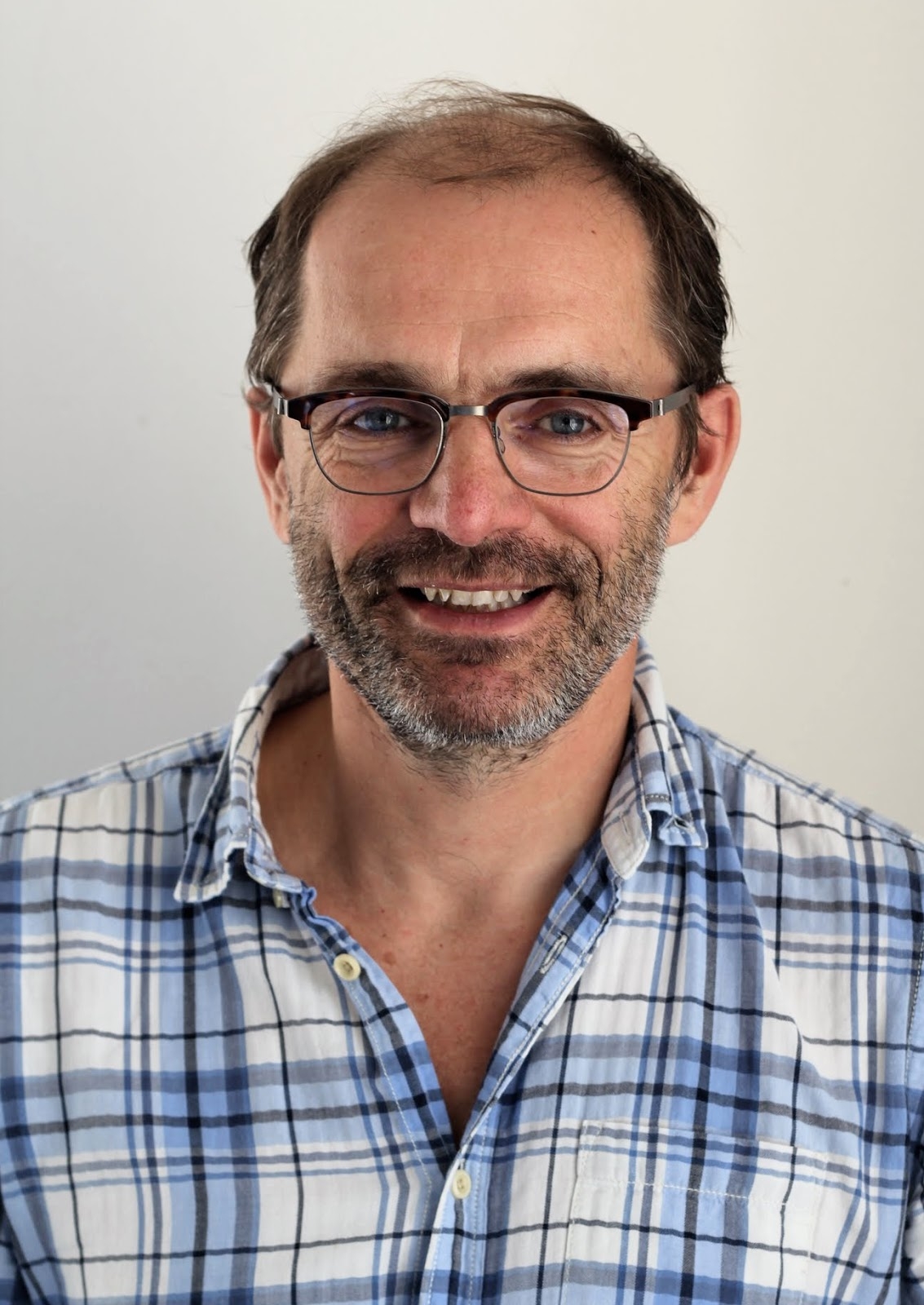}}]{Jiri Matas }
is the head of the Visual Recognition Group at the Department of Cybernetics, Czech Technical University in Prague. He holds a PhD degree from the University of Surrey, UK (1995). He has published more than 300 papers that have been cited about 75000 times in Google Scholar (h-index = 99).  He received the best paper prize at the British Machine Vision Conference in 2002, 2005 and 2022,  the Asian Conference on Computer Vision in 2007 and at the Int. Conf. on Document analysis and Recognition in 2015.   He is an Editor-in-Chief of the International Journal of Computer Vision was an Associate Editor-in-Chief of IEEE T.  Pattern Analysis and Machine Intelligence. He is on the computer science panel of the ERC.
\end{IEEEbiography}

\end{document}